\documentclass[paper]{JHEP3}
\usepackage{epsfig}
\usepackage{graphicx}
\usepackage{amsmath}
\usepackage{multirow}
\usepackage{lscape}
\usepackage{cite}

\usepackage{setspace}

\vspace{2cm}
\def\lsi{\raise0.3ex\hbox{$<$\kern-0.75em\raise-1.1ex\hbox{$\sim$}}}
\def\gsi{\raise0.3ex\hbox{$>$\kern-0.75em\raise-1.1ex\hbox{$\sim$}}}

\newcommand{\be}{\begin{equation}}
\newcommand{\ee}{\end{equation}}
\renewcommand{\d}{\textmd{d}}
\newcommand{\Tr}{\textmd{Tr}}

\usepackage{multirow}

\title{
The QCD phase diagram at nonzero quark density}
\author{
G.~Endr\H{o}di\footnote{Institute for Theoretical Physics, E\"otv\"os University, H-1117 Budapest, Hungary.} , Z. Fodor$^{*\dagger}$, S.D.~Katz$^{*\dagger}$, K.K. Szab\'o\footnote{Department of Physics, University of Wuppertal, D-42097 Wuppertal, Germany}
}

\abstract{
We determine the phase diagram of QCD on the $\mu-T$                                                                                       
plane for small to moderate chemical potentials. 
Two transition lines are defined with two quantities, 
the chiral condensate and the strange quark                                                                                     
number susceptibility. The calculations are                                                                                  
carried out on $N_t =6,8$ and $10$ lattices generated with a Symanzik                                                                                        
improved gauge and stout-link improved 2+1 flavor staggered fermion action                                                                                   
using physical quark masses. After carrying out the continuum extrapolation                                                                                  
we find that both quantities result in a similar curvature of the transition line.
Furthermore, our results indicate that in leading order 
the width of the transition region
remains essentially the same as the chemical potential is increased.
}


\begin{document}

\section{Introduction}

The understanding of the phase diagram of QCD is of utmost  importance and has attracted much attention, both experimental and theoretical. Experimental results are coming from cosmology and heavy ion collisions. Recently, in a collision of gold nuclei at the Relativistic Heavy Ion Collider (RHIC), a temperature beyond 200 MeV was reached~\cite{Adare:2009qk}, which indicates that the 
quark-gluon plasma has been created. Furthermore, the density of the system can be varied by tuning the
center of mass energy $\sqrt{s_{NN}}$.  
While most of the ongoing experiments like those at LHC or
RHIC concentrate  on achieving very high energies and thus small
chemical potentials, there are  projects that aim for regions of the
phase diagram with larger  densities (RHIC II, Facility for Antiproton and Ion Research (FAIR)). In these latter experiments an important objective is to identify the critical endpoint, e.g., by  searching for critical opalescence. Designing these next generation experiments can benefit greatly from developing theoretical understanding of the phase diagram.

\begin{figure}[ht!]
\centering
\mbox{
	\includegraphics*[height=6.2cm]{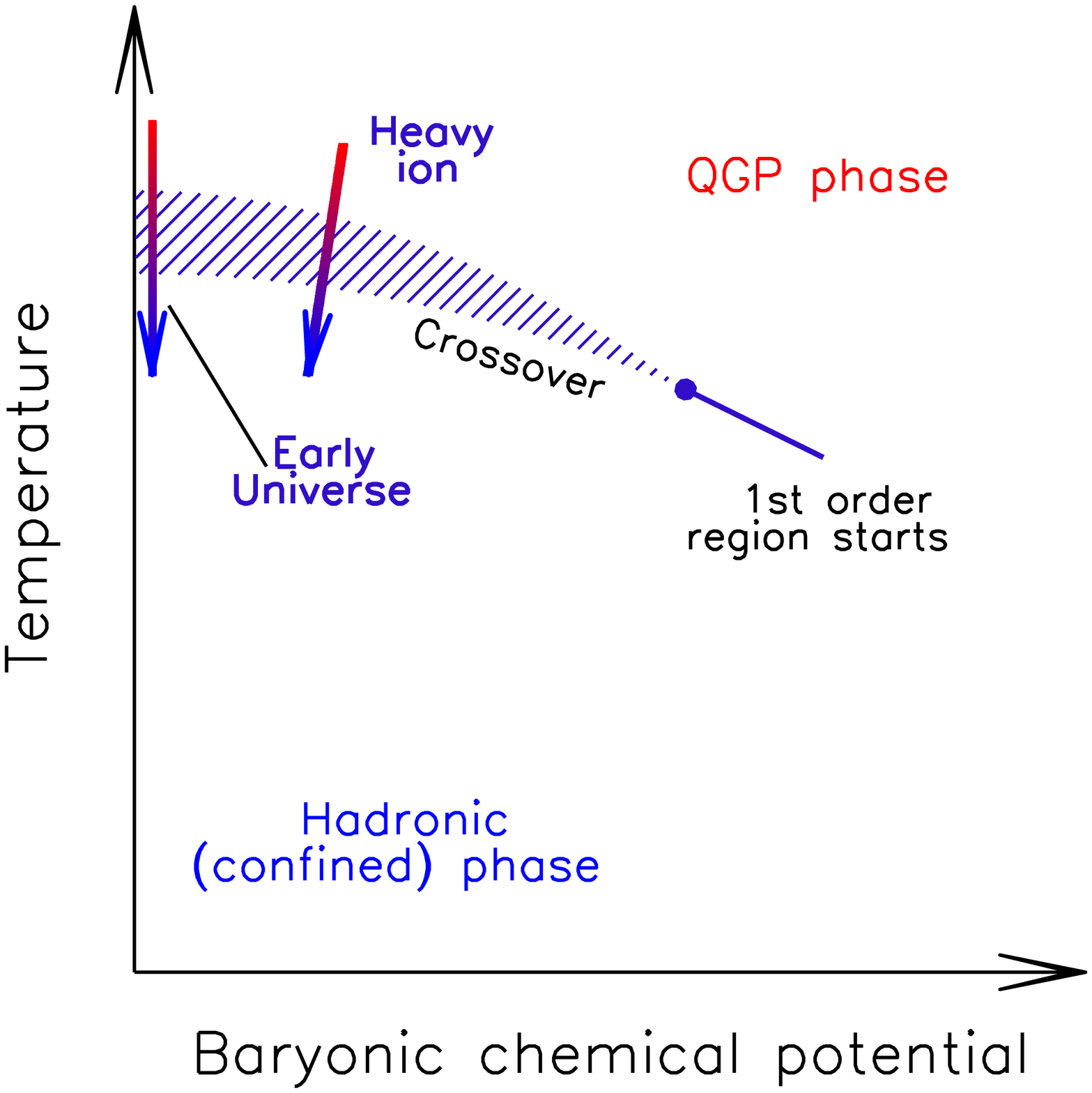}
	\includegraphics*[height=6.2cm]{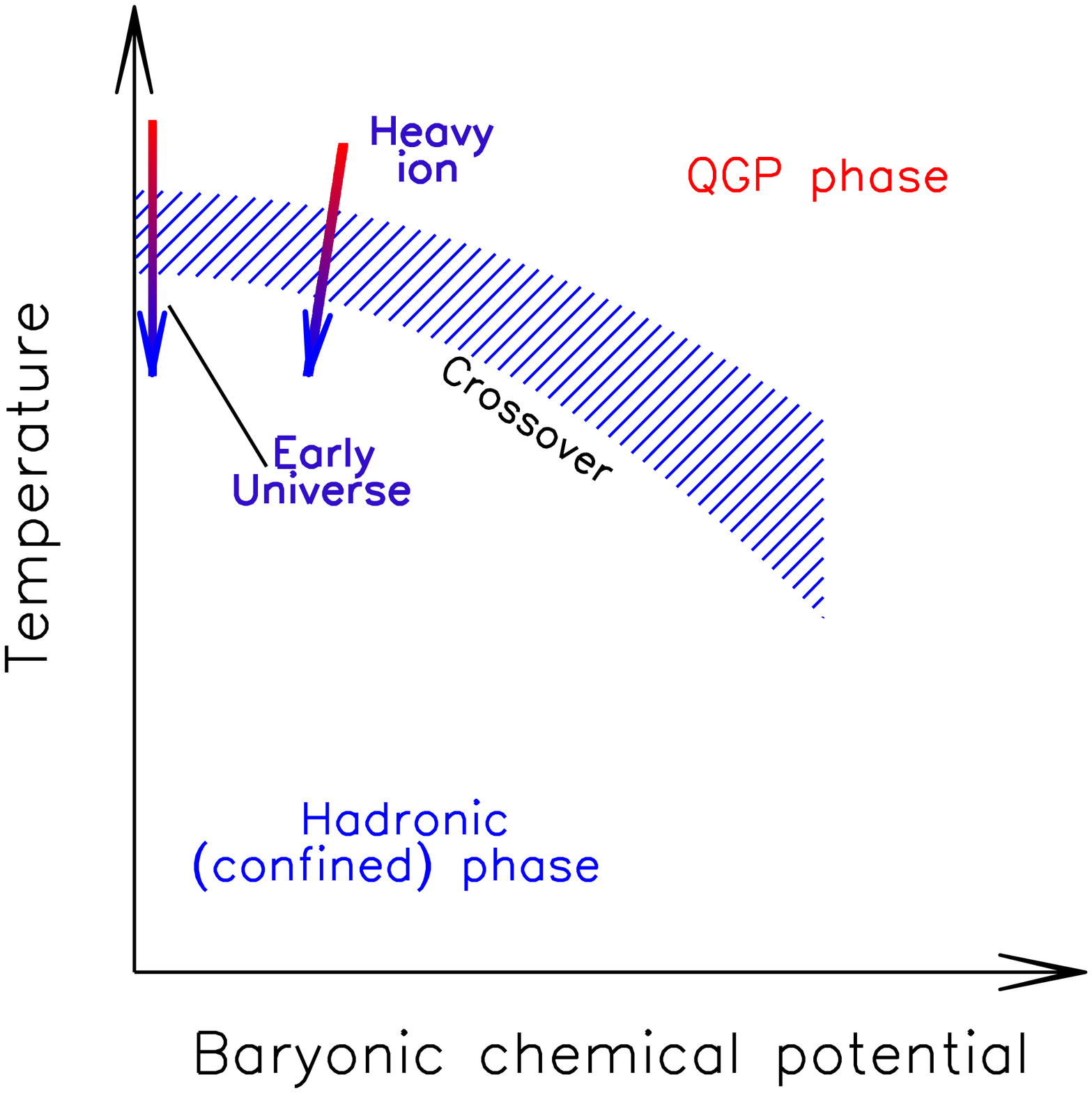}
}
\caption{
\label{fig:scenarios}
Two possible scenarios for the QCD phase diagram on the $\mu-T$ plane,
defined using a given observable. The left panel shows a phase diagram
with a transition growing stronger and possibly even 
turning into a real,
first-order phase transition at a critical endpoint. The right panel
on the other hand corresponds to a scenario with a weakening
transition and no critical endpoint. The paths corresponding to
systems describing the early Universe and a heavy ion collision are
also shown by the arrows. 
Note that different observables may lead to different scenarios. 
}
\end{figure}

Our theoretical knowledge about the phase diagram of QCD is mostly limited to the
zero chemical potential ($\mu=0$) axis and obtained by the use of
lattice QCD. The main reason that full results for $\mu>0$ are
not available is the infamous sign problem which spoils any lattice
technique based on importance sampling.
There are various
scenarios for the $\mu>0$ region of the phase diagram, among which 
two are illustrated in Figure~\ref{fig:scenarios}.

The transition at $\mu=0$ is a crossover~\cite{Aoki:2006we}
and we expect that the transition temperature decreases as we increase
$\mu$. Besides the actual value of the curvature of the transition line a particularly
interesting question is whether the transition becomes weaker or
stronger as $\mu$ grows. A strengthening of the
transition could lead to
the existence of a critical point, where the crossover transforms into a true phase transition (see left side of
Figure~\ref{fig:scenarios}). Another possibility is that the
transition weakens with increasing $\mu$ (see right side of
Figure~\ref{fig:scenarios}). 
The existence of the critical point would not be ruled out by such a scenario but would require non-monotonic behavior~\cite{Kapusta:2009ac}.

At zero
chemical potential lattice calculations provide reliable and accurate
results~\cite{Cheng:2006qk,Aoki:2006br,Aoki:2009sc,Bazavov:2009zn,Borsanyi:2010bp,Borsanyi:2010cj}.
Much more difficult is the situation at nonzero chemical potential.
Simulations at non-vanishing chemical potential are burdened by the
sign (complex action) problem: the fermion determinant here becomes
complex, and as a result makes Monte-Carlo methods based on importance
sampling impossible. Recently several methods were developed to access
the region of small chemical potentials. They are all based on
simulations at zero or purely imaginary chemical potentials where the
sign problem is absent. The first possibility is the reweighting of the
generated
configurations~\cite{Fodor:2001au,Fodor:2001pe,Csikor:2004ik,Fodor:2004nz, Fodor:2007vv}. The weight factors can also be approximated by a 
Taylor expansion in $\mu$~\cite{Allton:2002zi,Allton:2005gk,Gavai:2008zr,Basak:2009uv,Karsch:2010td}. Further possibilities are an analytic 
continuation from imaginary $\mu$~\cite{deForcrand:2002ci,D'Elia:2002gd,Wu:2006su,D'Elia:2007ke,Conradi:2007be,deForcrand:2008vr,D'Elia:2009tm,Moscicki:2009id}, or using the canonical ensemble~\cite{Alexandru:2005ix,Kratochvila:2005mk,Ejiri:2008xt}. 
The above studies were carried out on 
coarse lattices and in most cases with non-physical quark masses. 
We emphasize that to have a full result, the use of physical 
quark masses and a reliable continuum extrapolation are essential. 
In this paper we determine the transition temperature $T_c(\mu)$ as a function of the
chemical potential through a Taylor-expansion technique. The first term of this expansion 
is zero due to the symmetry $\mathcal{Z}(\mu)=\mathcal{Z}(-\mu)$ of the partition function. Therefore the
first nonvanishing contribution comes from the second order, which is related to the curvature
of the transition line.

\section{Definition of the curvature} 
\label{sec:def} 

Let us parameterize the transition line in the vicinity of the vertical $\mu = 0$ axis as
\be
T_c(\mu^2)=T_c \left( 1 - \kappa \cdot \mu^2/T_c^2 \right)
\ee
with $T_c$ being short for $T_c(0)$. This implies that the curvature can be written as 
\be
\kappa= - T_c \left. \frac{\d T_c(\mu^2)}{\d (\mu^2)} \right|_{\mu=0}
\label{eq:curvdef}
\ee
where (and also in the following) $\mu$ refers to the 
baryonic chemical potential ($\mu\equiv\mu_B=3\mu_{u,d}$), where $\mu_{u,d}$ is
the quark chemical potential assigned to the light quarks. Thus one has to measure $T_c$ as a function
of $\mu$ for small chemical potentials. To this end we use a definition of $T_c$ which is most suitable 
for determining the curvature.

Let us consider a quantity $\phi(T,\mu^2)$ that is monotonic in $T$ in the transition region, and fulfills the following 
constraints:
\be
\lim_{T\rightarrow 0} \frac{\partial}{\partial \mu^2} \phi(T,\mu^2) = 0, \quad\quad \lim_{T\rightarrow \infty} \frac{\partial}{\partial \mu^2} \phi(T,\mu^2) = 0 
\label{eq:constraints}
\ee
that is to say, $\phi$ does not depend on the chemical potential in 
the limiting cases $T\rightarrow 0$ and $T\rightarrow \infty$. For any fixed $\mu$ we can define a
transition temperature $T_c(\mu^2)$ as the temperature 
at which $\phi(T,\mu^2)$ takes the predefined constant 
value $C$:
\be
\left.\phi(T,\mu^2)\right|_{T=T_c(\mu^2)}=C.
\ee
We will choose a $C$ that corresponds to the inflection point 
of $\phi(T,0)$.
(Note that $T_c$ can also be defined as the location of the
maximum or inflection point of some observable. At non-zero $\mu$ this 
turns out to be somewhat less advantageous since a fitting of the reweighted data is 
required.)

Now let us determine the curvature using this definition of $T_c(\mu^2)$. 
The total derivative of the observable $\phi(T,\mu^2)$ may be written as
\be
\d\phi=\left. \left (\partial \phi/\partial T \right )\right|_{\mu=0} \cdot \d T + \left. \left (\partial \phi/\partial (\mu^2) \right )\right|_{\mu=0}\cdot \d \mu^2
\ee
Along the $T_c(\mu^2)$ line, $\phi$ is constant by definition, thus 
$\d\phi=0$. One obtains
\be
\frac{\d T_c}{\d \mu^2}=  \underbrace{ - \left.\left( \frac{\partial \phi}{\partial \mu^2} \right)\right|_{\substack{T=T_c\\\hspace*{-0.1cm}\mu=0}} \Big / \left.\left(\frac{\partial \phi}{\partial T} \right)\right|_{\substack{T=T_c\\\hspace*{-0.1cm}\mu=0}} }_{R(T)}
\label{eq:tcdef}
\ee
Thus, for every $C$ we can define a curvature. Since the $T_c(C)$ function is
invertible for the whole $C$ range, we can also write 
(\ref{eq:tcdef}) as a function of temperature, $R(T)$.

\begin{figure}[ht!]
\centering
\includegraphics*[height=6.0cm]{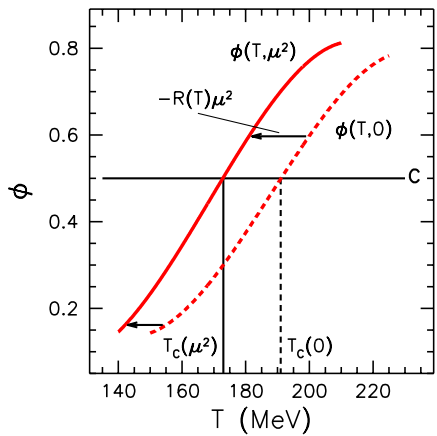}
\caption{Illustration of the behavior of the observable $\phi$ at $\mu=0$ and $\mu>0$. The quantity $T_c(\mu^2)$ is defined as the temperature where $\phi(T,\mu^2)$ crosses a constant value $C$. For $\mu>0$, each point of the $\phi(T,0)$ curve shifts in $T$ by $R(T)\cdot \mu^2$ (see definition in text).}
\label{fig:illustr}
\end{figure}

The function $R(T)$ is related to the distance that the $\phi(T)$ curve 
shifts along the $T$ axis as the chemical potential is varied. Given
$\phi(T)$ and $R(T)$ at zero chemical potential, the shift for non-zero
$\mu$ at leading order is $R(T) \cdot \mu^2$ (the curve moves to the left if $R(T)$ is negative and to
the right otherwise). This behavior is illustrated in Figure~\ref{fig:illustr}.
Using $R(T)$ we can define a temperature dependent curvature according to (\ref{eq:curvdef})
as $\kappa(T)=-T_c \cdot R(T)$. The meaning of $\kappa(T)$ is again simple:
it gives the curvature of the $\phi=\rm{const.}$ curve which starts from
$T$ at $\mu=0$.

We use the value of $\kappa(T)$ at $T=T_c$ to define the curvature
for a given observable. The shape of the $\kappa(T)$ function
also has important consequences. The slope of $\kappa(T)$ around $T_c$
is related to the width of the transition as follows:
if the slope is zero, i.e. $\kappa(T)$ is constant around $T_c$, then
all points shift the same amount along the $T$ axis when a small chemical potential
is switched on. This means that to leading order in $\mu$ the shape of
the $\phi(T)$ function (and thus, the width of the transition) does 
not change. If the slope is positive, then points with larger $T$ shift
more than the ones with smaller $T$ resulting in a compression of points, i.e. a 
narrower transition. Similarly, a negative slope indicates a broadening 
of the transition.

All in all, the expression $\partial \kappa / \partial T$ is therefore related to the relative change in the width $W(\mu)$ of the transition as the chemical potential increases:
\[
\frac{1}{W} \frac{\partial W}{\partial (\mu^2)} =- \left . \frac{1}{T_c}\frac{\partial \kappa}{\partial T} \right |_{T=T_c}
\]
where we assume that $W$ is proportional to the inverse slope of the quantity in question: $W \sim \left | \left.(\partial \phi / \partial T\right |_{T=T_c})^{-1} \right |$.

The two observables we use are the renormalized chiral condensate $\phi = \langle \bar\psi\psi_r\rangle$ and the normalized strange 
quark number susceptibility $\phi=\langle \chi_s/T^2 \rangle$. As we  show in section \ref{sec:observables},
both satisfy the constraints listed
in~(\ref{eq:constraints}).
The derivative $\partial \phi/\partial T$ is determined numerically, using the
$\mu=0$ data as a function of the temperature.
In order to calculate the derivative $\partial \phi/\partial (\mu^2)$
we need to measure more complicated operators; the technique for computing
these is detailed in the next section.

\section{Determination of the Taylor-coefficients}
\label{sec:tech}
Let us consider the partition function of the staggered lattice formulation for $N_f$ fermion flavors in
its usual form
\be
\mathcal{Z}=\int{\mathcal{D}U e^{-S_g(U)} (\det M)^{N_f/4}}
\ee
\noindent
and denote the derivative with respect to $\mu_{u,d}$ by $'$. The
derivatives of $\mathcal {Z}$ are easily calculated to be $(\log
\mathcal{Z})'=\langle n_{u,d} \rangle$ and $(\log
\mathcal{Z})''=\langle n_{u,d}^2 + n_{u,d}' \rangle - \langle n_{u,d}\rangle^2 $, where the light quark
number density $n_{u,d}$ and its derivative with respect to $\mu_{u,d}$ are given by the following combinations:
\begin{align*}
n_{u,d} &= \frac{N_f}{4}\Tr\left( M^{-1}M'\right)\\
n_{u,d}' &= \frac{N_f}{4} \Tr\left( M^{-1}M'' - M^{-1}M'M^{-1}M'\right)
\end{align*}
\noindent
Using these definitions the second derivative of any (possibly
explicitly $\mu$-dependent) observable can be straightforwardly determined. For the renormalized chiral condensate and the strange quark number susceptibility (see definition in section~\ref{sec:observables}) one obtains:
\begin{align}
\label{eq:pbpder}
\left.\frac{\partial^2 \langle \bar\psi\psi_r \rangle}{\partial \mu_{u,d}^2}\right|_{\mu_{u,d}=0} &=\langle \bar\psi\psi_r (n_{u,d}^2+n_{u,d}')\rangle - \langle \bar\psi\psi_r \rangle \langle n_{u,d}^2+n_{u,d}'\rangle +\langle 2\bar\psi\psi_r'n_{u,d} + \bar\psi\psi_r''\rangle \\
\left. \frac{\partial^2 \langle \chi_s \rangle}{\partial \mu_{u,d}^2} \right|_{\mu_{u,d}=0} &=\langle \chi_s (n_{u,d}^2+n_{u,d}')\rangle - \langle \chi_s \rangle \langle n_{u,d}^2+n_{u,d}'\rangle -2\langle n_s n_{u,d}\rangle^2
\label{eq:chisder}
\end{align}
\noindent
where $n_s$ is the strange quark number density, defined similarly as $n_{u,d}$. Note that the additive renormalization of $\bar\psi\psi_r$ (see section~\ref{sec:pbpdef}) does not influence the derivative in question.

For the chiral condensate -- being a $\mu$-dependent operator -- the derivatives $\bar\psi\psi_r'$ and $\bar\psi\psi_r''$ of this explicit dependence are also present in~(\ref{eq:pbpder}). These
terms were calculated numerically, using a purely imaginary chemical
potential $\Delta \mu_i$. The value of $\Delta \mu_i$ was varied in the
range $0.01 \ldots 0.0005$, and it was checked that the finite
differences converge fast enough to the $\Delta \mu_i \rightarrow 0$
values and the error coming from this approximation is negligible
compared to statistical errors. Taking into account these
considerations $\Delta \mu_i=0.001$ was used.

\section{The $\mu$-dependence of the observables}
\label{sec:observables}
We calculated the curvature of the transition line using the strange
quark number susceptibility and the chiral condensate. The details of
their renormalization and behavior are explained in this section.

\subsection{The strange quark number susceptibility}
The strange quark number susceptibility is defined as
\be
\langle \chi_s \rangle=\frac{T}{V}\frac{\partial^2 \log \mathcal{Z}}{\partial \mu_{s}^2}
\ee
This observable needs no renormalization, since it is connected to a
conserved current. It is useful to study
the combination $\langle\chi_s/T^2\rangle$, since it obeys the conditions
of~(\ref{eq:constraints}). It is easy to see that at $T=0$ one gets
$\langle\chi_s/T^2\rangle=0$ and at $T\to\infty$
the normalized quark number susceptibility 
$\langle\chi_s/T^2\rangle$ reaches its  $\mu_{u,d}$ independent 
Stefan-Boltzmann limit of 1.

\subsection{The chiral condensate}
\label{sec:pbpdef}
The chiral condensate can also be expressed as a derivative of the
partition function:
\be
\langle \bar\psi\psi \rangle=\frac{T}{V}\frac{\partial \log \mathcal{Z}}{\partial m}
\ee
The renormalization of $\bar\psi\psi$ is a more delicate issue as
compared to the situation with $\chi_s$. The free energy
($\log \mathcal{Z}$) contains additive divergences in the cutoff. 
In order to carry out the proper renormalization of the condensate, these
additive divergences have to be eliminated -- this is done by
subtracting the $T=0$ contribution.

The multiplicative divergence due to the derivative with respect to
the mass can be eliminated with a multiplication by the bare quark mass. 
Then, in order to have a dimensionless combination\footnote{Note that a 
division by $T^4$
which would also render the condensate dimensionless, changes the temperature dependence
and would lead to a non-monotonic $T$ dependence, which would be disadvantageous
in the present context.}, the whole expression can be divided by the fourth 
power of some dimensionful mass scale,  $Q^4$:
\be
\langle\bar\psi\psi_r\rangle=(\langle\bar\psi\psi\rangle - \langle\bar\psi\psi\rangle(T=0))\cdot m \cdot \frac{1}{Q^4}
\ee
This way no divergent contributions remain: this is a meaningful
quantity to study in the continuum limit.\footnote{Note that this renormalization
procedure leads to a somewhat unusual chiral condensate which vanishes
at $T=0$ and reaches a negative value at $T\to\infty$. A more conventional
condensate which is positive at $T=0$ and goes to zero at large temperatures
can be obtained by a constant shift which is irrelevant for our 
present study.}
In this work we use the $T=0$ pion
mass for the $Q$ normalization scale.

The final condition that has to be satisfied is that $\langle\bar\psi\psi_r\rangle$ 
should be independent of $\mu$ at $T=0$ and $T\to\infty$.
At $T=0$ the partition function is independent of $\mu$ as long as $\mu$ is
smaller than a $\mu_c$ critical value (the approximate baryon mass) 
and no baryons can be created from the vacuum.
Only for $\mu>\mu_c$ does the partition function have a non-trivial $\mu$ dependence.
Therefore all derivatives of $\mathcal{Z}$ (thus $\langle\bar\psi\psi_r\rangle$) 
are independent of $\mu$ for $\mu<\mu_c$. The chemical potential 
regime covered in this paper lies in this region. 
In the Stefan-Boltzmann limit ($T\to\infty$) the $\mu$                
independence is only satisfied in the sense 
$(\bar{\psi}\psi_r)^{-1} \partial/\partial \mu^2(\bar{\psi}\psi_r) \to 0$. 
Figure~\ref{fig:finV} demonstrates, however,        
that for temperatures above the transition 
region $\bar{\psi}\psi_r$ is already practically independent of $\mu$.

\begin{figure}[ht!]
\centerline{
\includegraphics*[height=5.6cm]{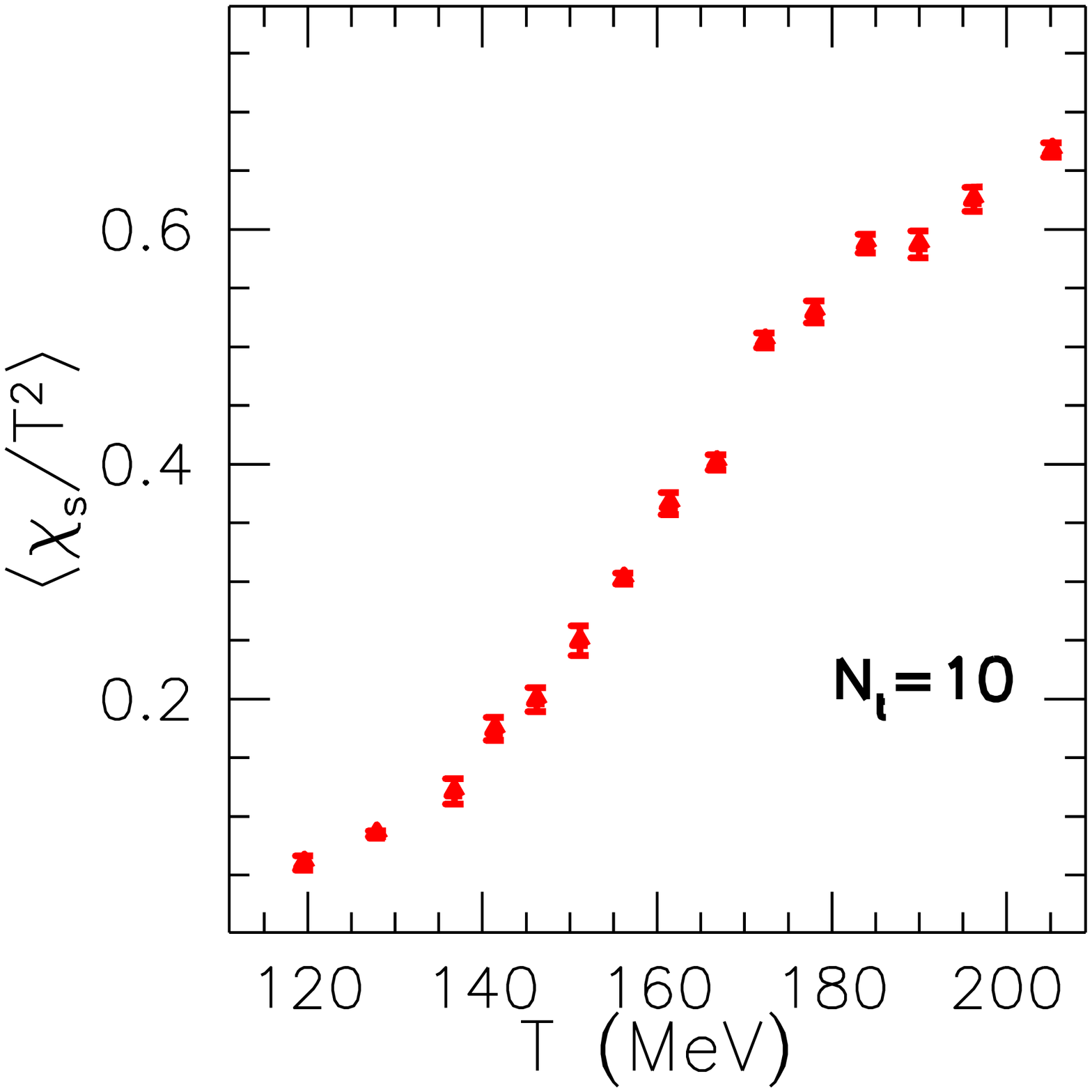}\hspace*{0.5cm}
\includegraphics*[height=5.6cm]{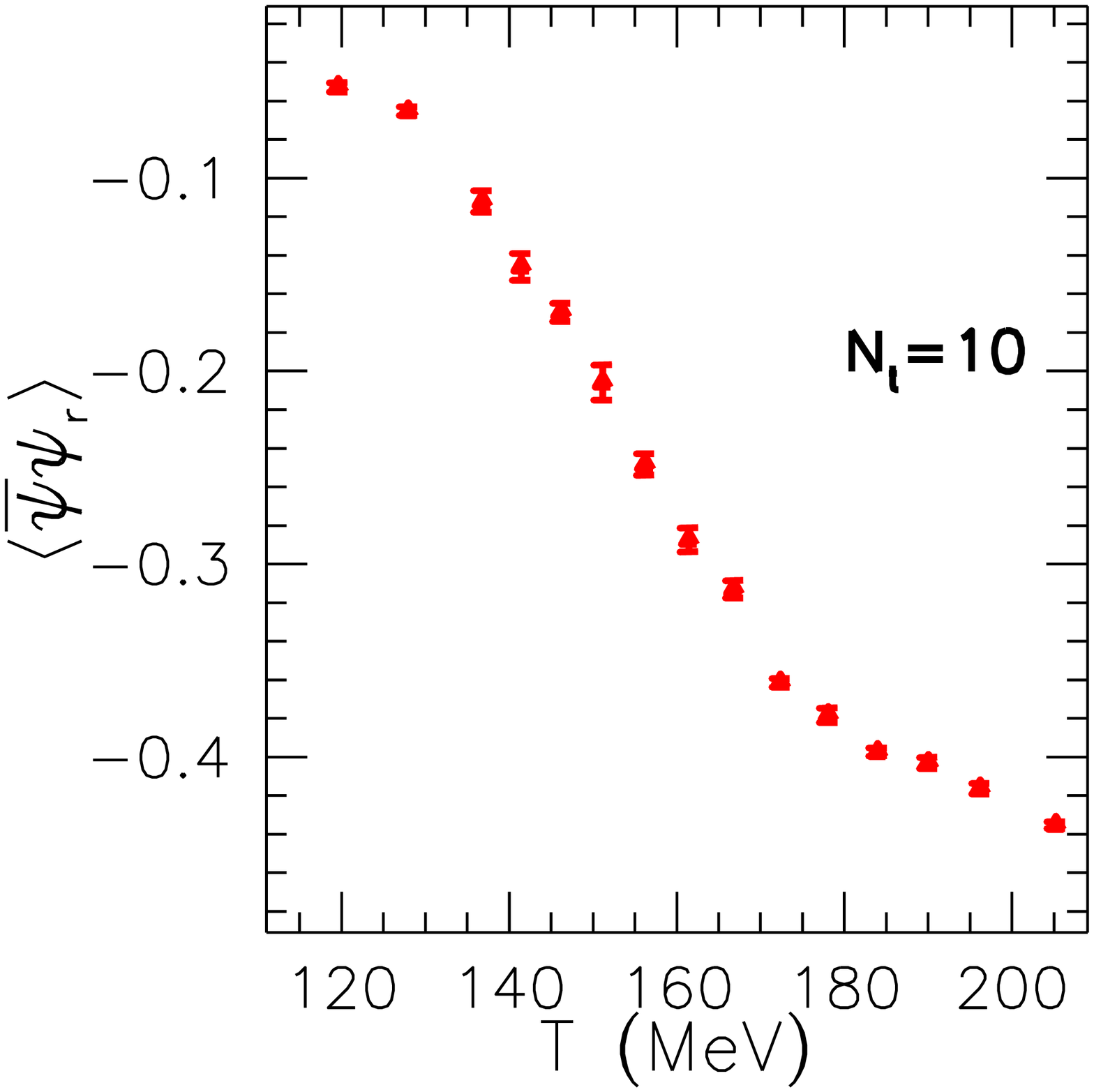}
}
\caption{The strange susceptibility and the renormalized chiral condensate as functions of the temperature at $\mu=0$.}
\label{fig:qsusc_pbp}
\end{figure}

Figure~\ref{fig:qsusc_pbp} illustrates the behavior of
$\langle\chi_s/T^2\rangle$ and $\langle\bar\psi\psi_r\rangle$ as a function of the temperature, 
determined on $N_t=10$ lattices.

\section{Simulation setup}
We used a Symanzik improved gauge and stout-link improved staggered
fermionic lattice action in order to reduce taste violation~\cite{Aoki:2006br}.
The configurations were generated with an exact RHMC algorithm~\cite{Clark:2006fx}. We
determined the line of constant physics (LCP) using physical masses for
the light quarks $m_{u,d}$ as well as for the strange quark $m_s$. The
LCP was fixed by setting the ratios $m_K/f_K$ and $m_K/m_\pi$ to their
physical values. We used three different lattice spacings $N_t = 6, 8,
10$ and aspect ratios $N_s/N_t$ of $4$ and $3$. The scale was fixed by
$f_K$ and its unambiguity was checked by calculating $m_{K*}$, $f_\pi$ and
$r_0$. The random noise estimator method was used to measure the operators
detailed in section~\ref{sec:tech}. We used 80 random vectors so that the
error coming from the method and the statistical error are of the
same extent. The details of the simulation setup can be found 
elsewhere~\cite{Aoki:2006br,Aoki:2005vt}.
We used the gauge ensembles
generated for a $\mu=0$ study~\cite{Aoki:2006br}. We also 
generated extra configurations for $N_t=8$ and 10. 
The number of trajectories at 
each $\beta$, $N_s$ and $N_t$ is summarized in table~1. 
The autocorrelation times
were below 10 trajectories in all cases.
After confirming the absence of thermalization effects,
we measured observables on every fifth trajectory. 
The measurements were performed on clusters equipped with graphics 
cards~\cite{Egri:2006zm}.

\begin{singlespace}
\begin{center}
\fontsize{8.3}{10.2}\selectfont
\mbox{
\begin{tabular}{| c | c | c | }
\hline
$N_s^3 \times N_t$ & $\beta$ & $\# \textmd{of trajecs.}$ \\ \hline
\multirow{10}{*}{$24^3 \times 6$} & 3.4500 & 1750 \\ \cline{2-3} 
& 3.4950 & 2500 \\ \cline{2-3} 
& 3.5100 & 5200 \\ \cline{2-3} 
& 3.5250 & 5350 \\ \cline{2-3} 
& 3.5400 & 5450 \\ \cline{2-3} 
& 3.5550 & 3400 \\ \cline{2-3} 
& 3.5700 & 3350 \\ \cline{2-3} 
& 3.5850 & 4650 \\ \cline{2-3} 
& 3.6000 & 3000 \\ \cline{2-3} 
& 3.6450 & 3650 \\ \hline
$18^3\times6$ & 3.5550 & 4550 \\ \hline
\multicolumn{3}{c}{}\\
\multicolumn{3}{c}{}\\
\multicolumn{3}{c}{}\\
\multicolumn{3}{c}{}\\
\multicolumn{3}{c}{}\\
\end{tabular}

\begin{tabular}{| c | c | c | }
\hline
$N_s^3 \times N_t$ & $\beta$ & $\# \textmd{of trajecs.}$ \\ \hline
\multirow{12}{*}{$24^3 \times 8$} & 3.6000 & 1800 \\ \cline{2-3} 
& 3.6250 & 2100 \\ \cline{2-3}
& 3.6375 & 4050 \\ \cline{2-3}
& 3.6500 & 5000 \\ \cline{2-3} 
& 3.6625 & 3150 \\ \cline{2-3} 
& 3.6750 & 3200 \\ \cline{2-3} 
& 3.6813 & 3200 \\ \cline{2-3} 
& 3.6875 & 14350 \\ \cline{2-3} 
& 3.7000 & 2100 \\ \cline{2-3} 
& 3.7160 & 3050 \\ \cline{2-3} 
& 3.7400 & 2850 \\ \hline
\multicolumn{3}{c}{}\\
\multicolumn{3}{c}{}\\
\multicolumn{3}{c}{}\\
\multicolumn{3}{c}{}\\
\multicolumn{3}{c}{}\\
\end{tabular}

\begin{tabular}{| c | c | c | }
\hline
$N_s^3 \times N_t$ & $\beta$ & $\# \textmd{of trajecs.}$ \\ \hline
\multirow{16}{*}{$28^3 \times 10$} & 3.6500 & 800 \\ \cline{2-3} 
& 3.6750 & 3350 \\ \cline{2-3} 
& 3.7000 & 800 \\ \cline{2-3} 
& 3.7125 & 850 \\ \cline{2-3} 
& 3.7250 & 800 \\ \cline{2-3} 
& 3.7375 & 800 \\ \cline{2-3} 
& 3.7500 & 2300 \\ \cline{2-3} 
& 3.7625 & 800 \\ \cline{2-3} 
& 3.7750 & 5950 \\ \cline{2-3} 
& 3.7875 & 4200 \\ \cline{2-3} 
& 3.8000 & 1600 \\ \cline{2-3} 
& 3.8125 & 2000 \\ \cline{2-3} 
& 3.8250 & 4850 \\ \cline{2-3} 
& 3.8375 & 4150 \\ \cline{2-3} 
& 3.8550 & 5950 \\ \hline 
\multicolumn{3}{c}{}\\
\end{tabular}

\label{tab:1}
}
\\
\normalsize
{\bf Table 1:} Number of trajectories for various lattice geometries.
\end{center}

\end{singlespace}

\section{Results}

\begin{figure}[ht!]
\centerline{
\includegraphics*[height=7cm]{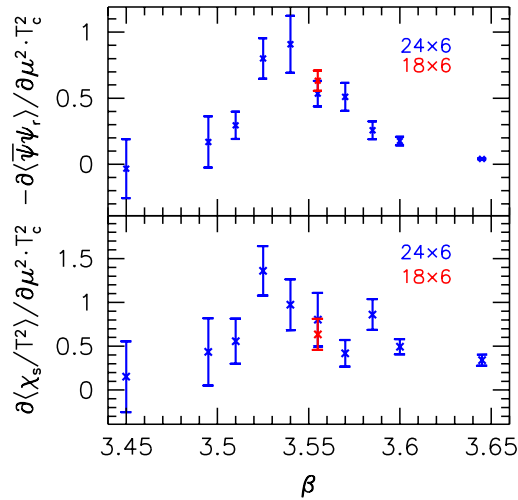}
}
\caption{The derivative of our observables with respect to $\mu^2$ measured on $N_t=6$ lattices. The $N_s=24$ results (blue points) are checked at one temperature by $N_s=18$ (red point). In the case of both observables a good agreement is observed, which indicates that finite size effects are small as compared to statistical errors.}
\label{fig:finV}
\end{figure}

First we checked finite size effects by comparing our results at $\beta=3.555$ obtained on $24^3\times6$ and on $18^3\times6$ lattices. This value of $\beta$ corresponds to about $155$ MeV, i.e. is near the pseudocritical temperature. The larger box is of physical size $\sim5$ fm. We observe a good agreement as the results for $\partial \phi/ \partial (\mu^2)$ agree within statistical errors for both the chiral condensate $\phi=\langle \bar\psi\psi_r \rangle$ and the strange quark number susceptibility $\phi=\langle\chi_s/T^2 \rangle$. Figure~\ref{fig:finV} shows our $N_t=6$ results for $N_s=24$ and $N_s=18$. Thus we conclude that finite size errors can be neglected at the present statistical accuracy.

\begin{figure}[ht!]
\centering
\mbox{
\includegraphics*[height=6.9cm]{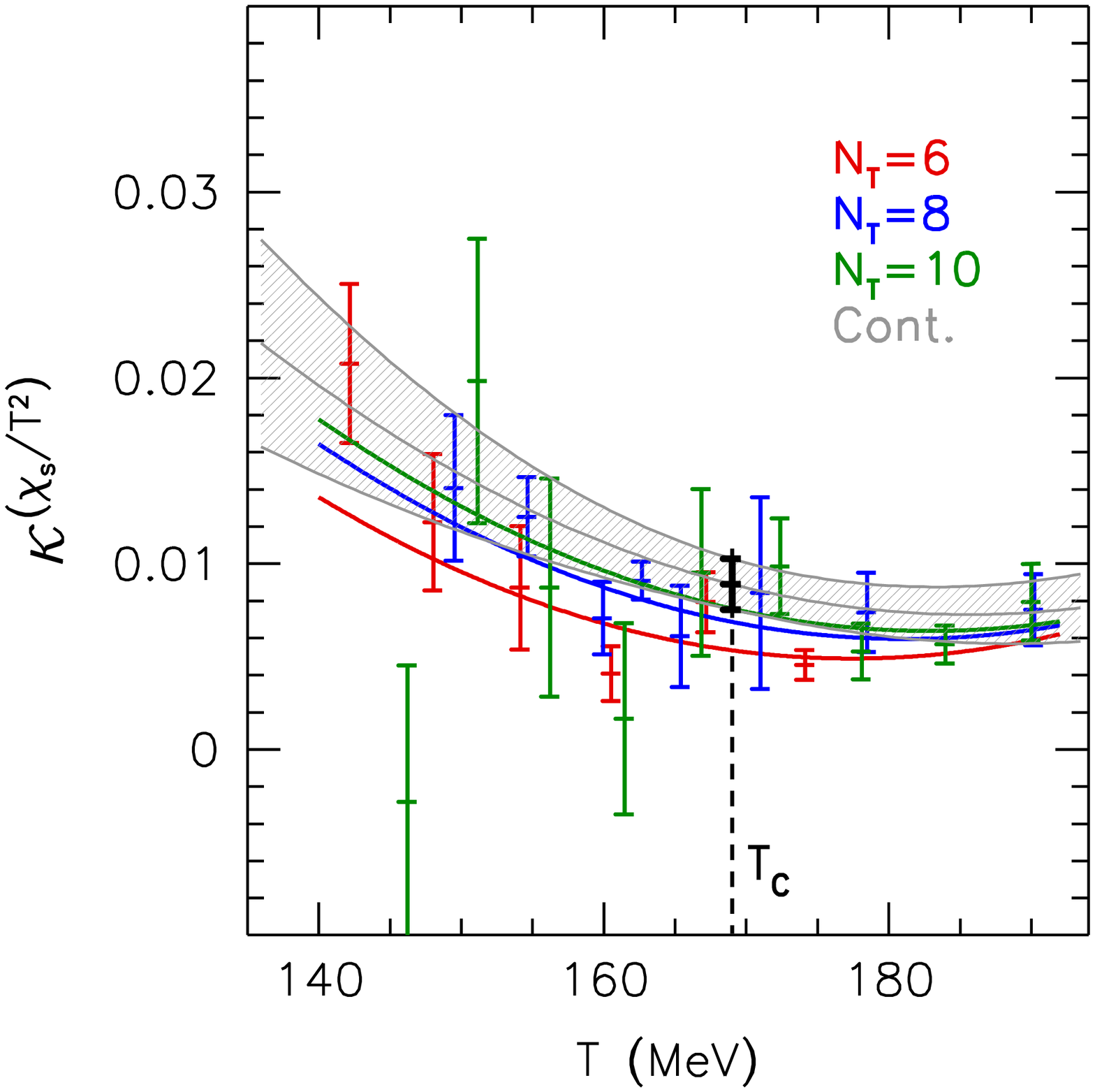}
\includegraphics*[height=6.9cm]{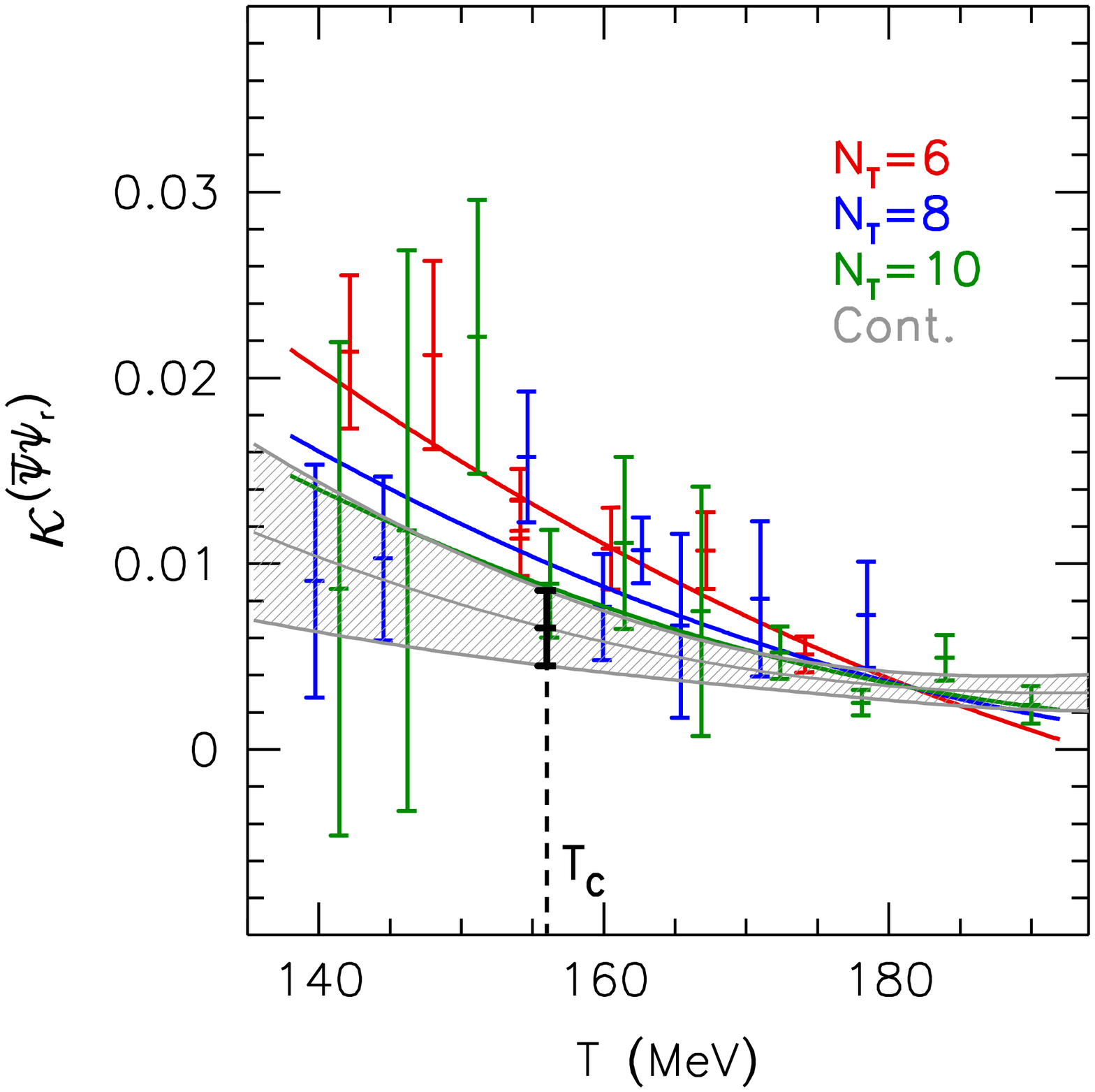}
}
\caption{
\label{fig:avg}
The curvature $\kappa(T)$ (see definition in text) determined using 
the strange quark number susceptibility (left) and the renormalized chiral 
condensate (right), respectively. A result of the combined fit (described in 
the text) is shown by the gray band. The fit results for the 
individual $N_t=6,8,10$ lattices are shown by the red, blue 
and green curves. The width of the gray band corresponds to the statistical
uncertainty of the fit.
}
\end{figure}

Since the actual shape of the $\kappa(T)$ function is unknown we carry
out a Taylor expansion around $T_c$ in the $t=(T-T_c)/T_c$
dimensionless variable:
\be
\kappa(T)=\kappa(T_c)+c_0\cdot t+c_1\cdot t^2
\ee
For each lattice spacing (i.e. each $N_t$) we have several simulation
points, corresponding to different temperatures.
In order to fit all
of our points at once, we allow a lattice spacing dependence for the
constant and linear terms (having a lattice spacing dependence
 of the quadratic term is also possible, but it does not improve the quality of the fits). Therefore we fit all of our simulation points with the following function
\be
\kappa(T;N_t)=\kappa(T_c;\textmd{cont})+c_0\cdot
t+c_1\cdot t^2+c_2/N_t^2+c_3\cdot t/N_t^2
\ee with fit parameters
$\kappa(T_c;\textmd{cont}),c_0,c_1,c_2$ and $c_3$.
The independent data points as well as the fitted curves (for each
$N_t$ and in the continuum) are shown in Figure~\ref{fig:avg}.

The $\chi^2/\textmd{d.o.f.}$ values of the two
 fits are 1.19 and 1.29,
respectively, indicating good fit qualities. The
 continuum curvatures
are given by the $\kappa(T_c;\textmd{cont})$ fit parameter, while the relative change
in the width of the transition can be read off from $-c_0$. Our final
results are 
\begin{align*}
\kappa^{(\chi_s/T^2)} &= 0.0089(14), & \kappa^{(\bar\psi\psi_r)} &= 0.0066(20) \\
\Delta W/W^{(\chi_s/T^2)}  &= 0.033(16), & \Delta W/W^{(\bar\psi\psi_r)} &= 0.030(18)
\label{eq:result}
\end{align*}
The results obtained from the two quantities are consistent with each other.
Using the $\kappa$ values we can give the transition lines
defined by any of the observables as
\be
T_c(\mu)=T_{c;\mu=0}[1-\kappa\cdot\mu^2/T_{c;\mu=0}^2]
\ee

The results for $\Delta W/W$ also suggest that the transition remains a weak crossover
with essentially constant strength 
for small to moderate chemical potentials.
Actually, there is a slight increase in the width of the 
transition determined from
both quantities. This effect is, however, very weak: the width only changes by a few percent 
up to $\mu\approx T_c$. 
This finding is consistent with previous results in the literature. In Ref.\cite{Fodor:2004nz} 
the imaginary parts of the Lee-Yang zeroes of the partition function were studied with exact
reweighting. At sufficiently high $\mu$ the Lee-Yang zeroes approached
the real axis, thus leading to a critical point. However, around $\mu=0$
an opposite effect can be observed, the Lee-Yang zeroes slightly 
move away from the real axis, indicating a weakening of the transition.
In Refs~\cite{deForcrand:2008vr} (3 flavors) and~\cite{Moscicki:2009id} 
(2+1 flavors) the critical surface in the quark masses -- 
chemical potential space was studied and the curvature of the surface 
suggests
a weakening of the transition as the chemical potential is increased.
Since all these leading order results predict a weakening of the transition for real chemical
potentials (i.e. $\mu^2>0$), a strengthening is expected in the 
imaginary direction ($\mu^2<0$). It is interesting to note that these 
leading order results are of the same order of magnitude in the sense that
using an extreme extrapolation they all lead to a critical point for an imaginary
chemical potential $\mu_q=i\cdot\mu_I$ in the range $\mu_I/T\approx 1-3$.

The validity range of our result is difficult to estimate from the present study alone. 
A conservative estimate
for the limit where the result obtained through the Taylor-expansion is still reliable
is $\mu_{u,d}\approx T_c$ i.e. where the expansion parameter exceeds unity. In the baryonic 
chemical potential this corresponds to about 500 MeV. Beyond this 
limit higher-order corrections are by all means expected to be important. 
Earlier experience with the exact reweighting method~\cite{Fodor:2004nz}
also shows that the leading-order quadratic behavior of the $T_c(\mu)$ function
dominates upto the above mentioned limit in the baryonic chemical potential.
To investigate whether higher order terms may lead to a critical point
one must carry out a similar analysis with full reweighting, 
beyond the reach of present computational resources. 

\begin{figure}[ht!]
\centering
\includegraphics*[height=9.0cm]{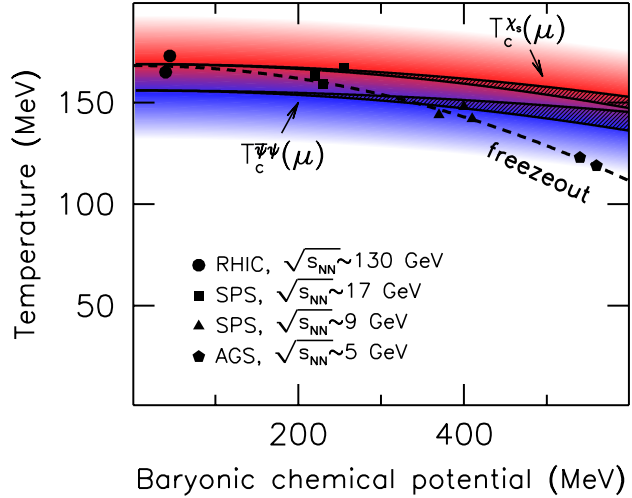}
\caption{
\label{fig:pd}
The crossover transition between the `cold' and `hot' phases is 
represented by the coloured area (blue and red correspond to the transition
regions obtained from the chiral condensate and the strange susceptiblity,
respectively). The lower solid band shows the result 
for $T_c(\mu)$ defined through the chiral condensate and the upper one 
through the strange susceptibility. The width of the bands represent the 
statistical uncertainty of $T_c(\mu)$ for the given $\mu$ coming from 
the error of the curvature $\kappa$ for both observables. The dashed line 
is the freeze-out curve 
from heavy ion experiments~\cite{Cleymans:1998fq}. Also indicated are with 
different symbols the individual 
measurements of the chemical
freeze-out from RHIC, SPS (Super Proton Synchrotron) and AGS 
(Alternating Gradient Synchrotron), respectively. The center of mass 
energies $\sqrt{s_{NN}}$ for each are shown in the legend.
}
\end{figure}

Our final result is shown in Figure~\ref{fig:pd}.
The crossover region's extent changes little as the chemical potential increases, and within 
it two definitions give different curves for $T_c(\mu)$. It is useful to
compare the whole picture to the freeze-out curve~\cite{Cleymans:1998fq} which summarizes 
experimental results on the $\{T,\mu\}$ points where hadronization of the quark-gluon plasma 
was observed. This curve is expected to lie in the interior of the crossover region, as is
indicated by our results as well.

\section{Acknowledgment} 
 We thank T. Cs\"org\H{o}, C. Schroeder and G.I. Veres for useful discussion. 
Computations were performed on the BlueGene at FZ J\"ulich and on clusters at Wuppertal and Budapest.
This work is supported in part by DFG grants SFB-TR 55, FO 502/1-2 and the EU 
grant (FP7/2007-2013)/ERC n$^o$208740.

\newpage

\bibliographystyle{JHEP}

\begin{thebibliography}{99}

\bibitem{Adare:2009qk}
  A.~Adare {\it et al.}  [PHENIX Collaboration],
  Phys.\ Rev.\  C {\bf 81}, 034911 (2010)
  [arXiv:0912.0244 [nucl-ex]].
\bibitem{Aoki:2006we}
  Y.~Aoki, G.~Endr\H{o}di, Z.~Fodor, S.~D.~Katz and K.~K.~Szab\'{o},
  Nature {\bf 443}, 675 (2006)
  [arXiv:hep-lat/0611014].
\bibitem{Kapusta:2009ac}
  J.~I.~Kapusta and E.~S.~Bowman,
  PoS C {\bf POD2009}, 018 (2009)
  [arXiv:0908.0726 [nucl-th]].
\bibitem{Cheng:2006qk}
  M.~Cheng {\it et al.},
  Phys.\ Rev.\  D {\bf 74}, 054507 (2006)
  [arXiv:hep-lat/0608013].
\bibitem{Aoki:2006br}
  Y.~Aoki, Z.~Fodor, S.~D.~Katz and K.~K.~Szab\'o,
  Phys.\ Lett.\  B {\bf 643}, 46 (2006)
  [arXiv:hep-lat/0609068].
\bibitem{Aoki:2009sc}
  Y.~Aoki, S.~Bors\'anyi, S.~Durr, Z.~Fodor, S.~D.~Katz, S.~Krieg and K.~K.~Szab\'o,
  JHEP {\bf 0906}, 088 (2009)
  [arXiv:0903.4155 [hep-lat]].
\bibitem{Bazavov:2009zn}
  A.~Bazavov {\it et al.},
  Phys.\ Rev.\  D {\bf 80}, 014504 (2009)
  [arXiv:0903.4379 [hep-lat]].
\bibitem{Borsanyi:2010bp}
  S.~Borsanyi, Z.~Fodor, C.~Hoelbling, S.~D.~Katz, S.~Krieg, C.~Ratti and K.~K.~Szabo
                  [Wuppertal-Budapest Collaboration],
  JHEP {\bf 1009}, 073 (2010)
  [arXiv:1005.3508 [hep-lat]].
\bibitem{Borsanyi:2010cj}
  S.~Borsanyi {\it et al.},
  JHEP {\bf 1011}, 077 (2010)
  [arXiv:1007.2580 [hep-lat]].
\bibitem{Fodor:2001au}
  Z.~Fodor and S.~D.~Katz,
  Phys.\ Lett.\  B {\bf 534}, 87 (2002)
  [arXiv:hep-lat/0104001].
\bibitem{Fodor:2001pe}
  Z.~Fodor and S.~D.~Katz,
  JHEP {\bf 0203}, 014 (2002)
  [arXiv:hep-lat/0106002].
\bibitem{Csikor:2004ik}
  F.~Csikor, G.~I.~Egri, Z.~Fodor, S.~D.~Katz, K.~K.~Szab\'o and A.~I.~T\'oth,
  JHEP {\bf 0405}, 046 (2004)
  [arXiv:hep-lat/0401016].
\bibitem{Fodor:2004nz}
  Z.~Fodor and S.~D.~Katz,
  JHEP {\bf 0404}, 050 (2004)
  [arXiv:hep-lat/0402006].
\bibitem{Fodor:2007vv}
  Z.~Fodor, S.~D.~Katz and C.~Schmidt,
  JHEP {\bf 0703}, 121 (2007)
  [arXiv:hep-lat/0701022].
\bibitem{Allton:2002zi}
  C.~R.~Allton {\it et al.},
  Phys.\ Rev.\  D {\bf 66}, 074507 (2002)
  [arXiv:hep-lat/0204010].
\bibitem{Allton:2005gk}
  C.~R.~Allton {\it et al.},
  Phys.\ Rev.\  D {\bf 71}, 054508 (2005)
  [arXiv:hep-lat/0501030].
\bibitem{Gavai:2008zr}
  R.~V.~Gavai and S.~Gupta,
  Phys.\ Rev.\  D {\bf 78}, 114503 (2008)
  [arXiv:0806.2233 [hep-lat]].
\bibitem{Basak:2009uv}
  S.~Basak {\it et al.}  [MILC Collaboration],
  PoS {\bf LATTICE2008}, 171 (2008)
  [arXiv:0910.0276 [hep-lat]].
\bibitem{Karsch:2010td}
  F.~Karsch {\it et al.},
  arXiv:1011.3130 [hep-lat].
\bibitem{deForcrand:2002ci}
  P.~de Forcrand and O.~Philipsen,
  Nucl.\ Phys.\  B {\bf 642}, 290 (2002)
  [arXiv:hep-lat/0205016].
\bibitem{D'Elia:2002gd}
  M.~D'Elia and M.~P.~Lombardo,
  Phys.\ Rev.\  D {\bf 67}, 014505 (2003)
  [arXiv:hep-lat/0209146].
\bibitem{Wu:2006su}
  L.~K.~Wu, X.~Q.~Luo and H.~S.~Chen,
  Phys.\ Rev.\  D {\bf 76}, 034505 (2007)
  [arXiv:hep-lat/0611035].
\bibitem{D'Elia:2007ke}
  M.~D'Elia, F.~Di Renzo and M.~P.~Lombardo,
  Phys.\ Rev.\  D {\bf 76}, 114509 (2007)
  [arXiv:0705.3814 [hep-lat]].
\cite{Conradi:2007be}
\bibitem{Conradi:2007be}
  S.~Conradi and M.~D'Elia,
  Phys.\ Rev.\  D {\bf 76}, 074501 (2007)
  [arXiv:0707.1987 [hep-lat]].
\bibitem{deForcrand:2008vr}
  P.~de Forcrand and O.~Philipsen,
  JHEP {\bf 0811}, 012 (2008)
  [arXiv:0808.1096 [hep-lat]].
\bibitem{D'Elia:2009tm}
  M.~D'Elia and F.~Sanfilippo,
  Phys.\ Rev.\  D {\bf 80}, 014502 (2009)
  [arXiv:0904.1400 [hep-lat]].
\bibitem{Moscicki:2009id}
  J.~T.~Moscicki, M.~Wos, M.~Lamanna, P.~de Forcrand and O.~Philipsen,
  Comput.\ Phys.\ Commun.\  {\bf 181}, 1715 (2010)
  [arXiv:0911.5682].
\bibitem{Alexandru:2005ix}
  A.~Alexandru, M.~Faber, I.~Horvath and K.~F.~Liu,
  Phys.\ Rev.\  D {\bf 72}, 114513 (2005)
  [arXiv:hep-lat/0507020].
\bibitem{Kratochvila:2005mk}
  S.~Kratochvila and P.~de Forcrand,
  PoS {\bf LAT2005}, 167 (2006)
  [arXiv:hep-lat/0509143].
\bibitem{Ejiri:2008xt}
  S.~Ejiri,
  Phys.\ Rev.\  D {\bf 78}, 074507 (2008)
  [arXiv:0804.3227 [hep-lat]].

\bibitem{Clark:2006fx}
  M.~A.~Clark and A.~D.~Kennedy,
  Phys.\ Rev.\ Lett.\  {\bf 98}, 051601 (2007)
  [arXiv:hep-lat/0608015].
\bibitem{Aoki:2005vt}
  Y.~Aoki, Z.~Fodor, S.~D.~Katz and K.~K.~Szab\'o,
  JHEP {\bf 0601}, 089 (2006)
  [arXiv:hep-lat/0510084].
\bibitem{Egri:2006zm}
  G.~I.~Egri, Z.~Fodor, C.~Hoelbling, S.~D.~Katz, D.~Nogr\'adi and K.~K.~Szab\'o,
  Comput.\ Phys.\ Commun.\  {\bf 177}, 631 (2007)
  [arXiv:hep-lat/0611022].
\bibitem{Cleymans:1998fq}
  J.~Cleymans and K.~Redlich,
  Phys.\ Rev.\ Lett.\  {\bf 81}, 5284 (1998)
  [arXiv:nucl-th/9808030].



\end{thebibliography}

\end{document}